# Gate-tunable Superconductivity in Hybrid InSb-Pb Nanowires

Yan Chen[1,*], David van Driel[2,*], Charalampos Lampadaris[1], Sabbir A. Khan[1,5], Khalifah Alattallah[1], Lunjie Zeng[3], Eva Olsson[3], Tom Dvir[2], Peter Krogstrup[4], Yu Liu[1,#]

[1]Center for Quantum Devices, Niels Bohr Institute, University of Copenhagen, 2100 Copenhagen, Denmark

[2]QuTech and Kavli Institute of NanoScience, Delft University of Technology, 2600 GA Delft, The Netherlands

[3]Department of Physics, Chalmers University of Technology, 41296, Gothenburg, Sweden

[4]NNF Quantum Computing Programme, Niels Bohr Institute, University of Copenhagen, 2100 Copenhagen, Denmark

[5]Danish Fundamental Metrology, 2970 Hørsholm, Denmark

We present a report on hybrid InSb-Pb nanowires that combine high spin-orbit coupling with a high critical field and a large superconducting gap. Material characterization indicates the Pb layer of high crystal quality on the nanowire side facets. Hard induced superconducting gaps and gate-tunable supercurrent are observed in the hybrid nanowires. These results showcase the promising potential of this material combination for a diverse range of applications in hybrid quantum transport devices.

Superconducting electronics is a rapidly evolving field with multiple potential applications. Such applications include parametric-amplifiers for qubit readout and control [1], and dissipation-less diodes [2,3]. Among candidate materials used for such electronic applications, the semiconductor-superconductor hybrids stand out, due to their ability to be tuned in-situ between various desirable regimes. The semiconductor-superconductor hybrids have also been proposed as a leading platform for the realization of Majorana zero modes, with potential application for fault-tolerant quantum computations. The elemental superconductor Al has so far proven to give highly transparent interface with semiconductors such as InAs or InSb, which has been demonstrated to be beneficial in various applications, such as superconducting diodes

[4], gatemon and fluxonium qubits [5, 6], parametric amplifiers [7], and "poor man's Majorana" states [8]. However, the small spectral gap and low critical field of Al might be limiting factors in pushing such applications forward. An alternative superconductor could be the elemental Pb, which has a significantly large gap and a high critical field. The Pb monolayers are known to exhibit strong Rashba spin–orbit coupling.[9, 10] The further study has revealed that the magnetic Co–Si clusters can drive a Pb monolayer into a topological superconducting state on Si(111).[11] The InAs-Pb hybrid island devices have shown correlated 2 electron transport, establishing the possibility to induce superconductivity from Pb to semiconducting platforms [12]. With CdTe shells as a tunnel barrier, it is also possible to grow epitaxial Pb layers on the side facets of InSb nanowires (NWs) [13]. Another system of interest is a hybrid between Pb and PbTe NWs, which are also used for Majorana-based research [14-16].

In this work we report the development of hybrid NWs composed of Pb and InSb. We grew InSb-Pb NWs using molecular beam epitaxy and in-situ metallic deposition. To avoid ex-situ etching while making Josephson junctions and tunnel junctions, a process that significantly reduces the quality and yield of the produced devices, we used a shadow technique to partially mask the deposition of Pb on some segments of the NWs. This approach allows us to obtain in-situ junctions. The structural characterization of the NWs shows the InSb core of single crystalline zinc-blende phase and the Pb layer of high crystal quality. However, the Pb layer is not epitaxial on the NW side facets. The transport characterization shows that the Pb layer has a superconducting transition temperature similar to that of bulk Pb. The spectrum of the hybrid system, measured using tunnel junctions, reveals a hard superconducting gap. Moreover, the observation of the oscillations of the supercurrent in the in-situ Josephson junctions under both in-plane and out-of-plane magnetic fields confirms the extent of the proximity effect from the Pb layer into the InSb core.

Stem-assistant InSb NWs were grown on pre-fabricated 1/4 2-inch undoped n-type (100) zinc-blende InAs wafers (Semiconductor Wafer Inc.) with the vapor-liquid-solid (VLS) method in III-V growth chamber of a solid-source Varian GEN-II MBE system with the background pressure below 1E−10 Torr, which is similar to our previous work [17-19]. The as-grown InSb NWs were transferred from the III-V growth chamber to the dedicated chamber for metal deposition after growth. Prior to the Pb deposition, the sample holder and the sample were left for cooldown and temperature stabilization for

~ 6 hours. The substrates were about 27° tilted towards the Pb source. The Pb layers were nominally deposited on two side facets of NWs with in-situ electron beam evaporation with a substrate temperature of -144 °C, as measured by a thermo-coupling back sensor. Calibration with a dummy wafer shows the surface temperature is actually around -20 °C. The average Pb deposition rate was 1.6 nm/s. The oxygen with 10 Torr partial pressure was used to in-situ oxidize the NW surface for 10 s, which helped to stabilize the Pb layers. As the last step, the load lock was vented with N2 (500 Torr), and the sample was placed in this environment for 5 min, allowing the sample temperature to raise above 0 °C before unloading. In this work, there are several batches of NWs, which have slight differences among each other. In structural characterization, the nominal 40 nm Pb was deposited on the NWs to increase image contrast and get better view of Pb layers and in-situ grown junctions. The NWs for studying Pb superconductivity have 20 nm Pb and 2 nm AlOx capping (the 2 nm Al layer was deposited with a growth rate of 0.06 nm/s with in-situ electron beam evaporation, while keeping the sample on the same sample stage so that the Al layers could be perfectly overlapping with the Pb layers. This is to avoid possible Pb dewetting during sample transferring and unloading). The Pb layer of NWs used for hard gap and supercurrent study is nominal 30 nm. The growth time of NWs used for hard gap and supercurrent study was 60 min while that of others was only 40 min.

The 30 kV field emission scanning electron microscope (SEM) JEOL 7800F was used to investigate how the NWs grew in the trenches and to locate the NWs for device fabrication. During device fabrication, the locations of junctions were also rechecked with SEM after NW transfer. The FEl Titan 80-300 transmission electron microscope (TEM) was used at 300 kV for scanning transmission electron microscopy (STEM) imaging, electron diffraction and energy-dispersive X-ray spectroscopy (EDS). The microscope is equipped with a Schottky field emission gun (FEG), a CEOS probe Cs corrector as well as STEM annular dark field (ADF) detectors.

The system for bulk Pb superconductivity characterization is a modified Dynacool cryo-cooled physical property measurement system (PPMS) with a temperature range of 1.7 K – 400 K and perpendicular magnetic field in a range of 9 T. When measuring the differential conductance, the lock-in amplifiers (SR830) were employed to enable the PPMS to measure tiny AC signal. The SIS device presented in Fig. 2 was measured in a Janis cryostat with a vector magnet and at a base temperature of 300 mK. The standard lock-in amplifier techniques were used to obtain the differential conductance.

The SNS device of Fig. 3 was measured in an Oxford Triton dilution refrigerator with a base temperature of approximately 30 mK. Each contact was routed to two bondpads to perform four-point measurements. For each measurement, two of the four bondpads were used to apply a current bias while the other two were used to measure the voltage drop. The DC differential resistance was obtained using a numerical derivative of the VI-curves. All devices were tuned using a global back gate.

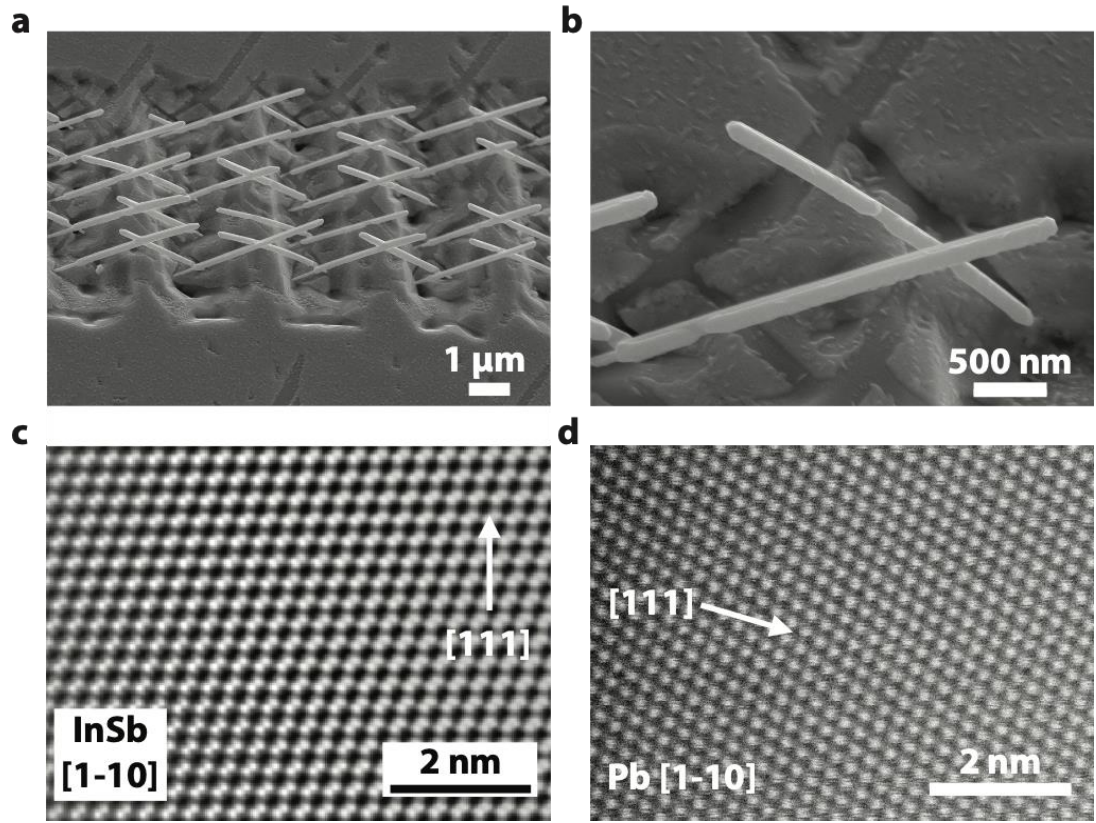

**Fig. 1.** Structure of InSb-Pb NWs. **a.** SEM image (viewing angle 36° from normal) of InSb-Pb NWs. **b.** Enlarged SEM image (viewing angle 36° from normal) of InSb-Pb NWs to show their junctions. **c.** Atomic resolution STEM ADF image of InSb lattice in InSb-Pb NWs. **d.** Atomic resolution STEM ADF image of Pb lattice in InSb-Pb NWs.

As shown in Fig. 1, the as-grown InSb NWs are typically about 200 nm of diameter and around 3.5 μm of length. The Pb layers with a nominal thickness of 40 nm were deposited on two side facets of InSb NWs. The in-situ grown junctions are visible in Fig. 1b, where the Pb beam is blocked by other NWs during deposition. The remaining part of the Pb layers is complete and continuous. There are small features on the surface of the Pb layers along NWs, which suggests that the substrate temperature is not low enough to reach the range suitable for Pb deposition. To understand the lattice match between InSb and Pb, we performed STEM ADF imaging on the InSb-Pb NWs at atomic resolution. In Fig. 1c, which shows a transverse view of the NW, we can see the single crystalline zinc-blende InSb lattice without dislocation when the zone axis is along InSb [1-10]. Under the same zone axis, the lattice fringes in the Pb shell are not visible, and the InSb/Pb interface is not sharp at the atomic scale (Fig. S1). Based on cross-sectional EDS in Fig. S2, it seems that part of InSb moves from the interface between InSb and Pb to the surface of the Pb shell, so the interface is planarized. Therefore, it is probably needed to introduce barrier layers like CdTe to protect InSb [13]. In comparison, both the Pb and InAs lattices are visible when zone axis is along InAs [11-20] on the InAs stem of the NWs (Fig. S3). It shows the Pb layer can grow epitaxially on InAs, which is consistent with the previous report [12]. We then rotated the NW and aligned the Pb crystal orientation with the electron beam. It is found that the

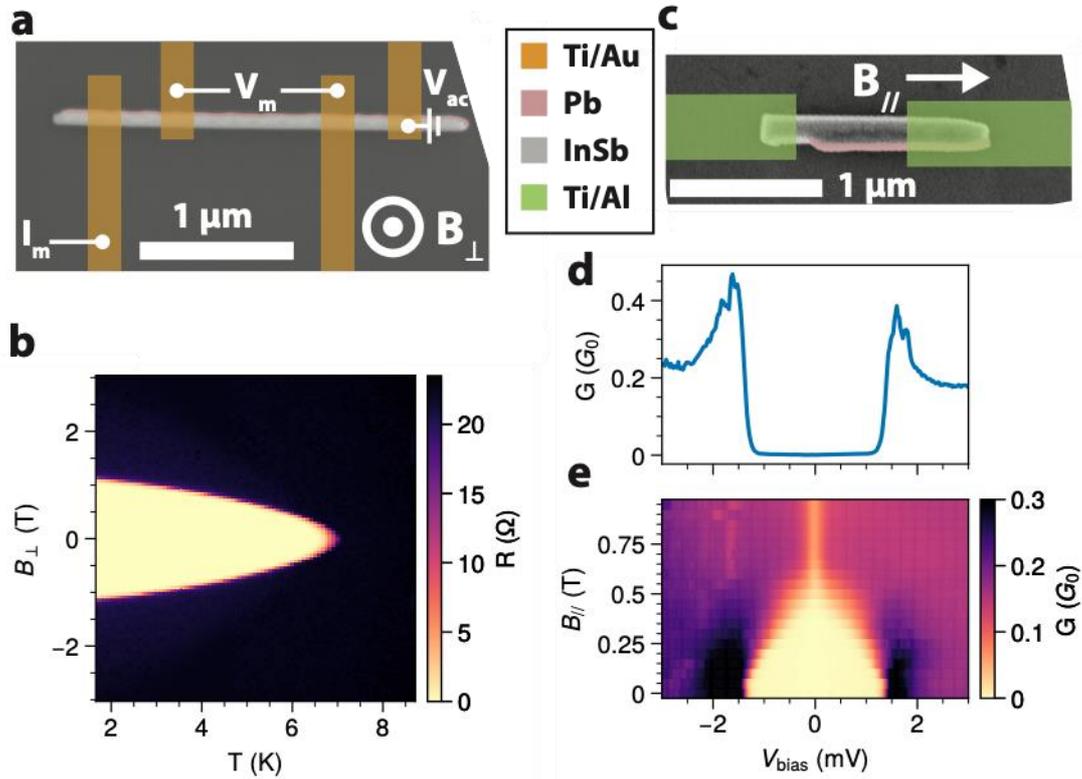

**Fig. 2 transport measurements of InSb-Pb hybrid NWs. a.** False-colored SEM micrograph of an InSb-Pb NW without junctions, where the contacts are schematically indicated by golden rectangles. **b.** Four-point measurements of zero-bias differential resistance as a function of out-of-plane magnetic field $B_\perp$ and temperature $T$ of the device presented in panel a. **c.** False-colored SEM micrograph of an InSb-Pb NW SIS device with Ti/Al contacts and a global back gate, where a magnetic field $B_{//}$ is applied parallel to the NW axis. **d.** Tunneling spectroscopy $G = dI/dV$ of the SIS device of panel c at $B = 0$ for varying voltage bias $V_{bias}$ measured at T = 300 mK. An AC voltage $V_{AC} = 30\,\mu V$ is applied on top of the voltage bias. **e.** Same as panel d, for varying both $B_{//}$ and $V_{bias}$.

Pb [1-10] zone axis is tilted 5~10° with respect to the InSb [1-10] zone axis based on analysis of 5 NWs, so the Pb layer is not epitaxial on InSb side facets. The STEM imaging (Figure 1d) shows the crystalline structure of Pb shell. The Pb layers usually prefer to grow with (111) surface.[12] In this case, the surface of Pb layer is a few degrees off from (111), so the lattice mismatch at the interface between InSb and Pb can probably be reduced. Meanwhile, the diffraction patterns acquired along the Pb layer suggest its single crystallinity, as shown in Fig. S4.

After growth of InSb-Pb NWs, we used the NWs without junction to study the properties of the Pb shell. Figure 2a shows an InSb-Pb NW contacted with four normal Ti/Au contacts. Based on the device, the 4-probe differential resistance of the Pb shell on top of the NW as a function of the temperature ($T$) and magnetic field applied in the out-of-plane direction ($B_\perp$) was measured as shown in Fig. 2b. An AC voltage excitation from a lockin amplifier was applied to one of the outer leads. The resulting current was measured at the other outer lead by a lockin, while the voltage drop between the two inner leads was measured by another lockin. The differential resistance was then computed by dividing the measured voltage drop by the measured current. No DC current or voltage bias was applied to the device. We observed a region of zero resistance, which turned into finite upon reaching a critical value of $B_\perp$. Moreover, this critical field decreases with increasing temperature, which shows characteristics of type-I superconductors [20]. The critical temperature $T_C$ is ~ 7.0 K, which is similar to the corresponding value for bulk Pb [21]. We extrapolate a zero-temperature critical field of ~1.3 T from the empirical, parabolic law for finite temperature critical fields $B_c(T) = B_c^0(1 - T^2/T_c^2)$ [20]. This is much larger than the bulk Pb value of $B_c^0 = 0.08T$ at 0 K [21]. Figure S5 shows similar four-point measurements performed on two other InSb-Pb NWs, with comparable values for $T_c$, but having different critical fields due to different orientations with respect to the magnetic field.

After having focused on the intact Pb layer, we now turn ourselves to InSb-Pb NWs with junctions. Figure 2c shows an InSb-Pb NW with an in-situ grown junction [18], contacted by 10 nm Ti + 50 nm Al leads on the bare InSb and the Pb respectively. We set $V_{bg} = 0.1\ V$, so the bare InSb acts as a tunnel barrier and perform superconductor–insulator–superconductor (SIS) spectroscopy for $B_{//} = 0$ as shown in Fig. 2d. We observed a single pair of particle-hole symmetric peaks at $eV_{bias} \approx \pm 1.6 meV$, instead of two pairs at $eV_{bias} = \pm(\Delta_{Al} \pm \Delta_{Pb})$ as expected from SIS tunneling. The relative magnitude of these peaks depends on temperature, and the peak at $\Delta_{Al} + \Delta_{Pb}$ has been observed to be more visible than the $\Delta_{Al} - \Delta_{Pb}$ peak in the previous study [22]. Taking the bulk value of $\Delta_{Al} = 0.175\ meV$ for the Al lead [23], we estimate $\Delta_{Pb} \approx 1.4\ meV$, similar to that of a Pb layer 1.36 meV [24] but larger than theoretical value 1.28 meV [25] or 1.25 meV of epitaxial Pb [12]. Moreover, the sub-gap conductance is two orders of magnitude lower than the above-gap conductance, which we interpret as a hard gap, similar to those reported in III/V NWs with epitaxial Al [26, 27], epitaxial Pb [12], polycrystalline Sn [28], and amorphous Ta [29]. Figure 2e shows tunneling spectroscopy for varying $B_{//}$, where the Al turns normal for a finite field. It is found that the gap shrinks with field but cannot be closed due to blockaded transport at low-bias. It can be attributed to an accidental quantum dot formed in the junction which causes a coulomb blockade.

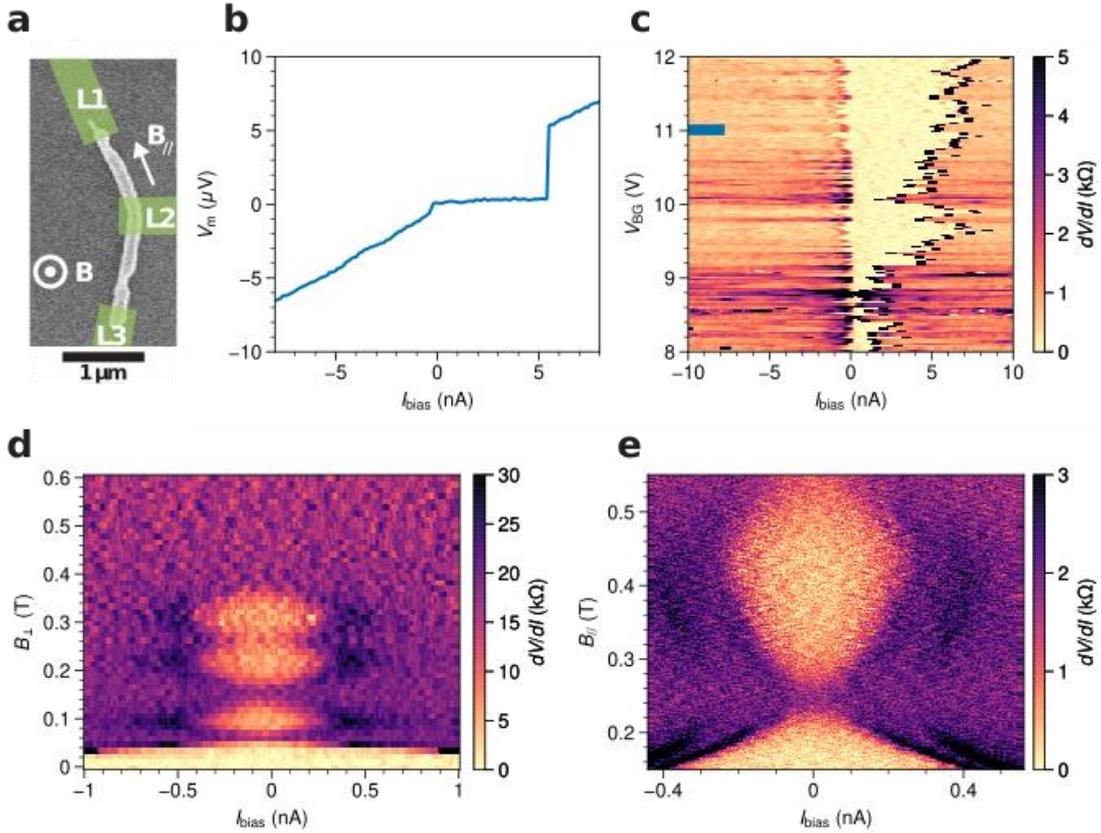

**Fig. 3. Current-biased measurements of InSb-Pb SNS devices. a.** False-colored SEM micrograph of an InSb-Pb NW SNS device, contacted on the Pb shell by three Ti/Al contacts. A global back gate is used to tune the InSb electron density. The directions of magnetic fields are shown for panels d and e. **b.** The voltage drop $V_m$ measured between contacts L2 and L3 indicated in panel a. **c.** The DC differential resistance $R_{diff} = dV_m/dI$ versus bias current $I_{bias}$, for varying back gate voltage $V_{BG}$. The blue line indicates the value of $V_{BG}$ where panel b is measured. **d.** $R_{diff}$ versus bias current $I_{bias}$ and out-of-plane magnetic field $B_\perp$, measured between contacts L1 and L2. **e.** Same as panel d, but for magnetic field $B_{//}$ applied parallel to the NW axis.

Another metric to characterize the proximity effect in semiconductor-superconductor hybrids is presence of supercurrents in Josephson junctions. In Fig. 3a, the InSb-Pb NW was shadowed in two segments, so two superconductor – normal conductor – superconductor (SNS) junctions were formed. Figure 3b shows the *V-I* curve measured in a four-point configuration between the middle and top contact depicted in Fig. 3a. Within a range of applied currents, between approx. 0 and 6 nA, no voltage dropped over the junction, indicating the presence of a supercurrent in the system. In Fig. 3c, the measured switching current ($I_S$) responds non-monotonically to $V_{BG}$, typical for a disordered Josephson junction in such hybrids. When we turn our attention to the dependence of $I_S$ on the applied magnetic field, Figure 3d shows *V-I* curves measured with increasing $B_\perp$ between the middle and bottom contact shown in Fig. 3a. $I_S$ decayed rapidly once the magnetic field was applied, giving rise to an interference pattern with $I_S$ smaller than 500 pA. The separation between the lobes allows us to estimate the effective area of the junction by assuming that they are separated by approximately one flux quantum. The lobes are spaced 100 mT apart, giving an effective junction are of 20,000 nm², slightly larger than the area of 15,000 nm² observed in the SEM image. In Fig. 3e, we show the interference pattern of $I_S$ resulting from the application of in-plane magnetic field. Here, there is a supercurrent extinction and revival around $B_{//} \approx 250 mT$, which corresponds roughly to one superconducting flux being threaded through the NW cross-section. This behavior is predicted for NW Josephson junctions with multiple modes [30, 31]. The observed gate-, and field-based interference could result from mode-mixing of multiple, occupied modes in the NW as observed in comparable systems [32, 33]. The maximum value of switching current is usually proportional to the superconducting gap [34]. Therefore, mixing of multiple modes in a large-gap superconductor like Pb could result in a gate-tunable supercurrent over a large range. It is noted that the NW in Fig. 3a is bent. When the NW is bent, the strain can be induced in the NW. The strain can move the Fermi level in the band structure, so it can influence the band offset between InSb and Pb. At the same time, it can change the carrier density which can affect the strength of proximity effect. Therefore, the induced superconducting gap will be changed, and the obtained results could be deviated from intrinsic properties of the devices based on strain-free NWs.

In summary, we develop hybrid InSb-Pb NWs based on molecular beam epitaxy growth technique. The in-situ junctions were fabricated during Pb deposition with a shadowing approach. The NWs consist of the single crystalline zinc-blende InSb lattice free of dislocations and the Pb layer of high crystal quality. However, the Pb [1-10] zone axis is tilted 5 ~10° with respect to the InSb [1-10] zone axis, so the Pb layer is not epitaxial on InSb side facets. The intact Pb layer shows a superconducting transition at 7 K, similar to that of bulk Pb. The tunneling spectrum of a SIS device shows a superconducting gap of around 1.40 meV in the single-particle density of states of the InSb-Pb NW. The sub-gap conductance is two orders of magnitude lower than that above the gap, indicating a hard gap. The supercurrent interference measurements show modulation of switching current for both in-plane and out-of-plane magnetic fields.

**AUTHOR INFORMATION**
*YC and DvD contribute equally to this work.
**Corresponding Author**
#E-mail: yu.liu@nbi.ku.dk.
**Note**
The authors declare no competing financial interests.

**ACKNOWLEDGMENTS**
We thank C. B. Sørensen, A. J. Cui, J. H. Kang for technical assistance. We also thank L. Kouwenhoven for fruitful discussion. We acknowledge financial support from the Microsoft Quantum initiative, from the Danish Agency for Science and Innovation through DANSCATT, from the European Research Council under the European Union's Horizon 2020 research and innovation program (grant agreement n° 716655), and from the international training network 'INDEED' (grant agreement n° 722176). This project has received funding from the European Union's Horizon 2020 research and innovation program under grant agreement No.823717-ESTEEM3. SAK also acknowledge financial support from Danish agency for higher education and science.

Supplementary information

Gate-tunable Superconductivity in Hybrid InSb-Pb Nanowires

Yan Chen[1,*], David van Driel[2,*], Charalampos Lampadaris[1], Sabbir A. Khan[1,5], Khalifah Alattallah[1], Lunjie Zeng[3], Eva Olsson[3], Tom Dvir[2], Peter Krogstrup[4], Yu Liu[1,#]

[1]Center for Quantum Devices, Niels Bohr Institute, University of Copenhagen, 2100 Copenhagen, Denmark

[2]QuTech and Kavli Institute of NanoScience, Delft University of Technology, 2600 GA Delft, The Netherlands

[3]Department of Physics, Chalmers University of Technology, 41296, Gothenburg, Sweden

[4]NNF Quantum Computing Programme, Niels Bohr Institute, University of Copenhagen, 2100 Copenhagen, Denmark

[5]Danish Fundamental Metrology, 2970 Hørsholm, Denmark

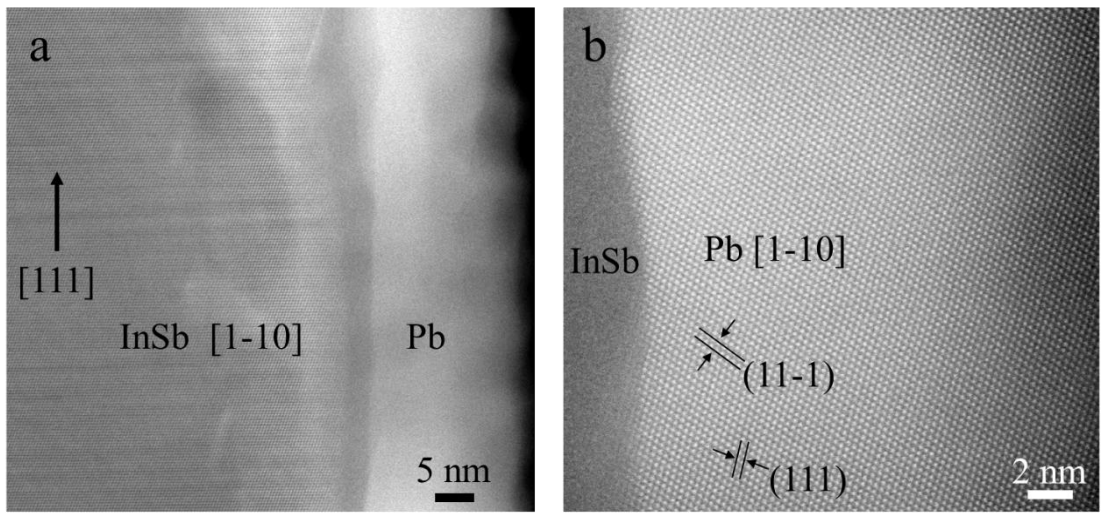

Figure S1. Interface of InSb-Pb NWs. (a) STEM image at InSb/Pb interface to show there were no lattice fringes in the Pb area when the zone axis is along InSb [1-10]. (b) STEM image at InSb/Pb interface when the zone axis is along Pb [1-10].

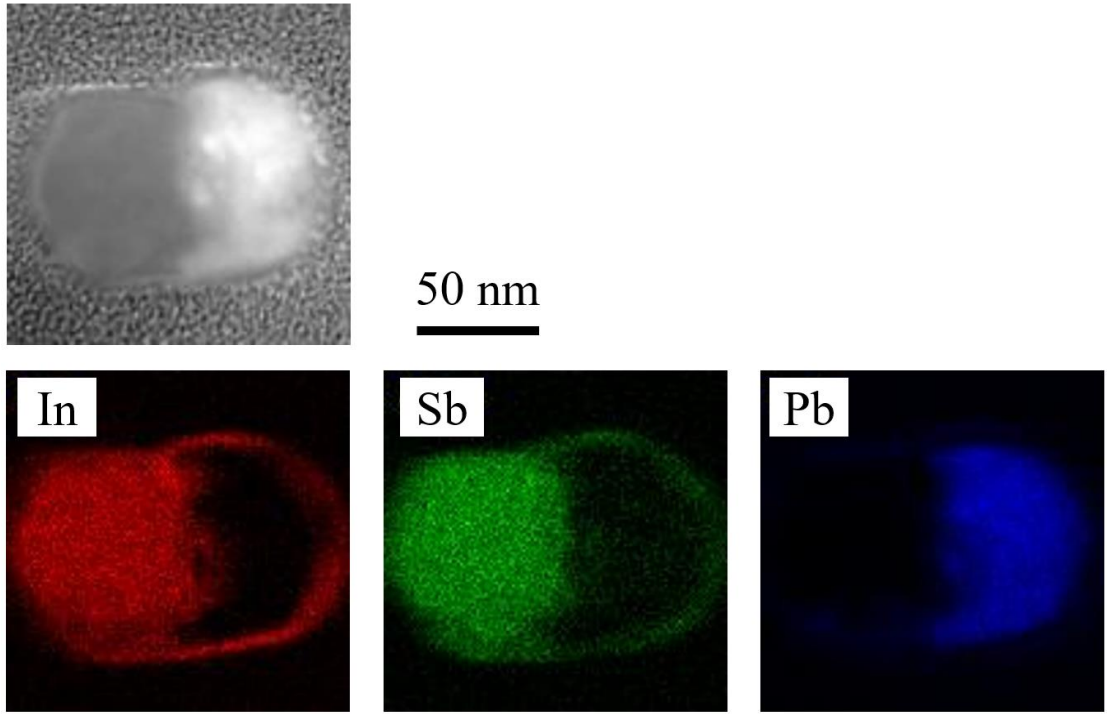

Figure S2. Cross-sectional energy-dispersive X-ray spectroscopy (EDS) of InSb-Pb NW. The signal from In and Sb is basically overlapping with each other, representing InSb, while the signal of Pb comes from the Pb shell. Part of InSb at the interface between InSb and Pb seems to dissolve and segregate to the surface of the Pb shell.

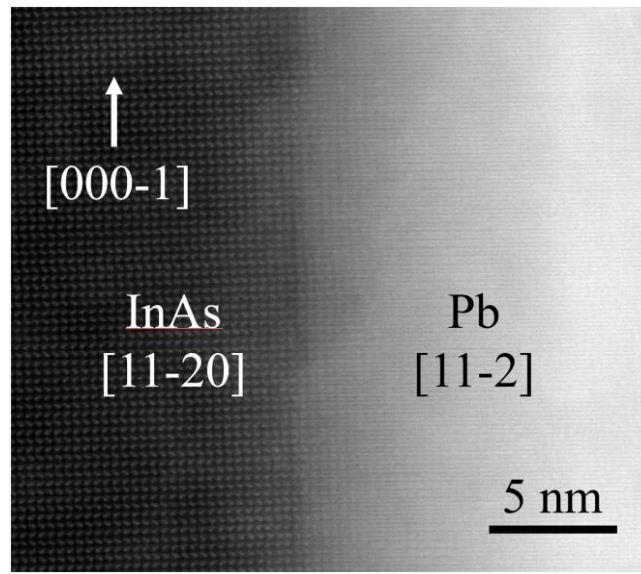

Figure S3. Atomic scale structure (STEM ADF image) of the InAs stem part with a Pb layer in InSb-Pb NWs.

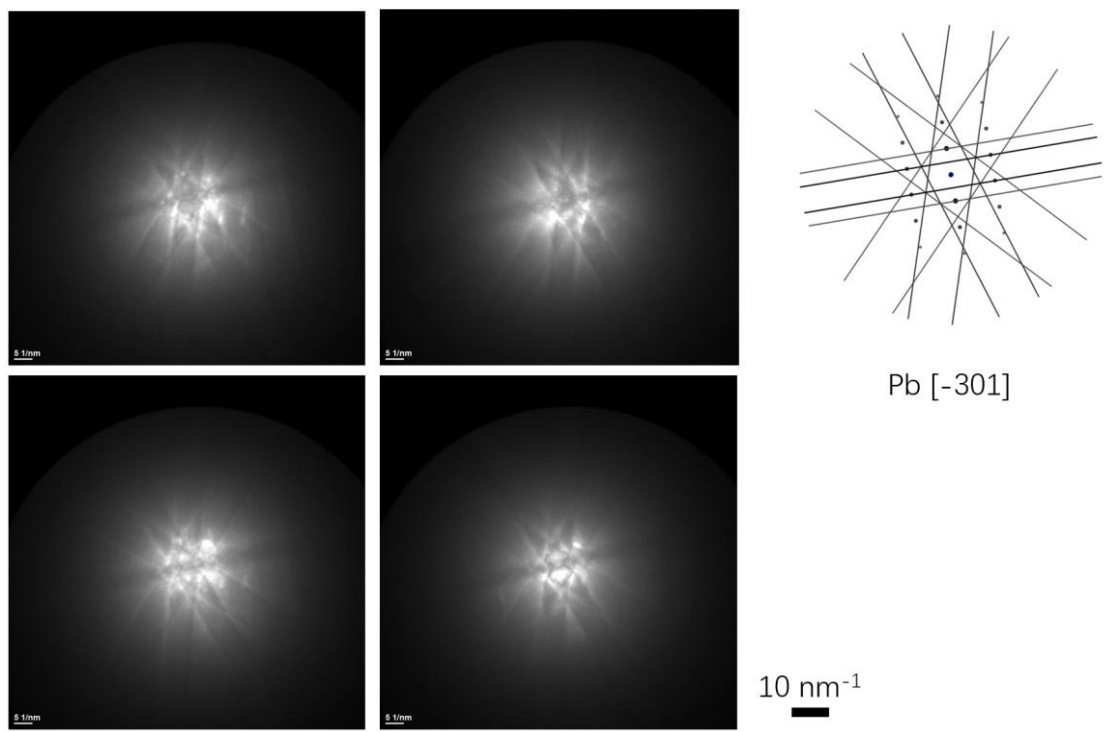

Figure S4. Future information about Pb layers of InSb-Pb NWs. Here, electron beam was put on different positions in Pb shell of one single NW. The diffraction pattern does not change with position, which means the structure and orientation of Pb does not change in the shell. It indicates the Pb shell is single-crystalline. The measurements were further performed on 3 NWs. They all show the same result. Based on the simulated Kikuchi pattern shown on the right, the Pb zone axis is indexed as [-301].

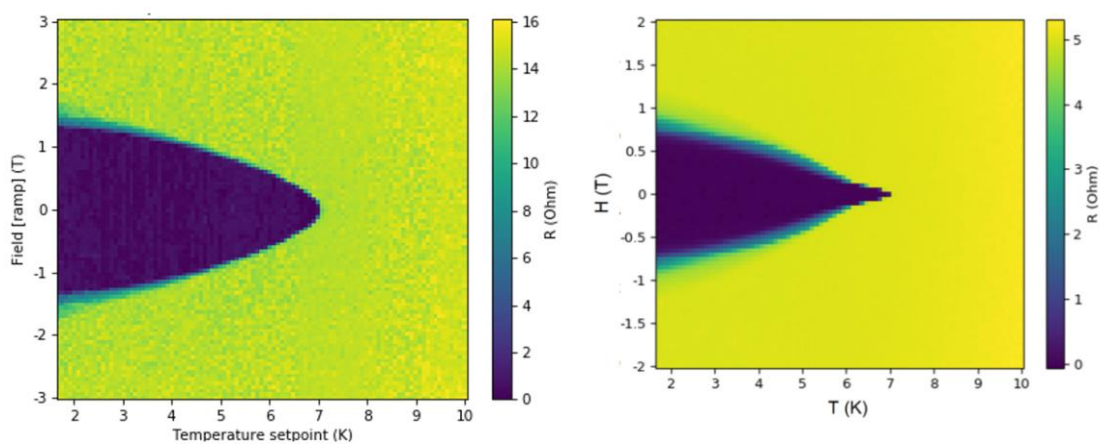

Figure S5. Superconductivity of Pb layer in other measured InSb-Pb NWs. The NWs have almost the same Tc. However, the geometries of NWs relative to the magnetic field are different, so they exhibit different Hc.